# Gravitational and Anti-gravitational Applications

*by*

*Roger Ellman*

Abstract

It is now possible to partially deflect gravitation away from an object so that the gravitational attraction on the object is reduced. That effect makes it possible to extract energy from the gravitational field, which makes the generation of *gravito-electric* power technologically feasible. Such plants would be similar to hydro-electric plants and would have their advantages of not needing fuel and not polluting the environment.

Physically, the action of deflecting away gravitational attraction, which of course is directed toward the gravitation source, produces an equal but opposite reaction on the mechanism that produces the deflection of the gravitational action [the deflector], a reaction directed away from the gravitation source.

The result is the combination of reducing the gravitational attractive acceleration of the object toward the gravitation source plus the introducing of a reactive acceleration on the object in the direction away from the gravitation source.

Such a deflector, engineered to enable controlled adjustment of the amount and the direction of its action, could provide for a spacecraft both launch levitation and deep space travel acceleration. It could provide both levitation and horizontal motion for a flying vehicle over a planet surface.

This technology, which uses readily abundantly available materials and techniques, is ready now for research and engineering refinement.

[Patent Pending *(P)*, January 13, 2011, USPTO #13/199,867.]

Roger Ellman, The-Origin Foundation, Inc.
    320 Gemma Circle, Santa Rosa, CA 95404, USA
    RogerEllman@The-Origin.org
    http://www.The-Origin.org

# *Gravitational and Anti-gravitational Applications*

by

*Roger Ellman*

Abstract

It is now possible to partially deflect gravitation away from an object so that the gravitational attraction on the object is reduced. That effect makes it possible to extract energy from the gravitational field, which makes the generation of *gravito-electric* power technologically feasible. Such plants would be similar to hydro-electric plants and would have their advantages of not needing fuel and not polluting the environment.

Physically, the action of deflecting away gravitational attraction, which of course is directed <u>toward</u> the gravitation source, produces an equal but opposite reaction on the mechanism that produces the deflection of the gravitational action [the deflector], a reaction directed <u>away</u> from the gravitation source.

The result is the combination of reducing the gravitational attractive acceleration of the object toward the gravitation source plus the introducing of a reactive acceleration on the object in the direction away from the gravitation source.

For example, an object experiencing a natural gravitational acceleration, $A$, reduced $80\%$ by gravitation deflection to $0.2 \cdot A$, plus simultaneously experiencing the reaction to the $80\%$ deflection in the amount $0.8 \cdot A$, experiences a net acceleration acting in the direction <u>away</u> from the gravitation source of $0.8 \cdot A - 0.2 \cdot A = 0.6 \cdot A$.

Of course, the $A$ is the Newtonian gravitational acceleration $G \cdot M / d^2$ where $M$ and $d$ are the mass of and distance to the gravitating source, for example the Sun, the Earth, or Mars.

Such a deflector, engineered to enable controlled adjustment of the amount and the direction of its action, could provide for a spacecraft both launch levitation and deep space travel acceleration. It could provide both levitation and horizontal motion for a flying vehicle over a planet surface.

Just as the sail-driven ships of past centuries experienced fuel-free travel by means of controlled manipulation of the energy of the wind, this technology enables fuel-free travel through space by means of controlled manipulation of the energy of the gravitational field that permeates all of space.

This technology, which uses readily abundantly available materials and techniques, is ready now for research and engineering refinement.

### SUMMARY DEVELOPMENT

Light normally travels in a straight direction. But, when some effect slows a portion of the light wave front the direction of the light is deflected. In Figure 1 below, the shaded area propagates the arriving light at a slower velocity, $v'$, than the original velocity, $v$, [its index of refraction, $n'$, is greater] so that the direction of the wave front is deflected from its original direction.

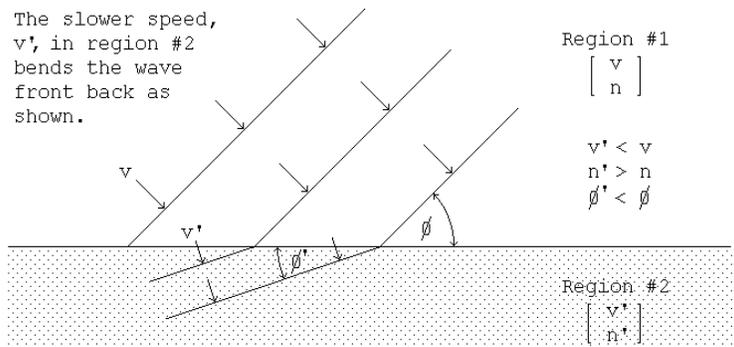

*Figure 1 - Deflection of Light's Direction
by Slowing of Part of Its Wave Front*

A slowing of part of its wave front is the mechanism of all bending or deflecting of light. In an optical lens, as in Figure 2 below, light propagates more slowly in the lens material than outside the lens. The amount of slowing in different parts of the lens depends on the thickness of the lens at each part. In the figure the light passing through the center of the lens is slowed more than that passing near the edges. The result is the curving of the light wave front.

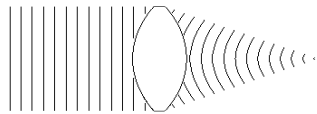

*Figure 2 - The Bending of Light's Wave Front by an Optical Lens*

"Gravitational lensing", shown below, is an astronomically observed effect in which light from a cosmic object too far distant to be directly observed from Earth becomes observable because a large cosmic mass [the "lens"], located between Earth observers and that distant object, deflects the light from the distant object as if focusing it, somewhat concentrating its light toward Earth enough for it to be observed from Earth. The light rays are so bent because the lensing object slows more the portion of the wave front that is nearer to it than it slows the farther away portion of the wave front.

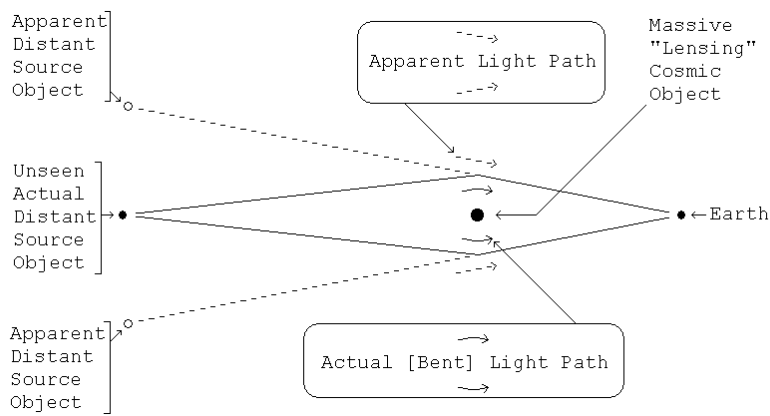

*Figure 3 - Gravitational Lensing Bending of Light Rays*

The same effect occurs on a much smaller scale in the diffraction of light at the two edges of a slit cut in a flat thin piece of opaque material as shown below. The bending is greater near the edges of the slit because the slowing is greater there. The effect of the denser material in which the slit is cut slows the portion of the wave front that is nearer to it more than the portion of the wave front in the middle of the slit.

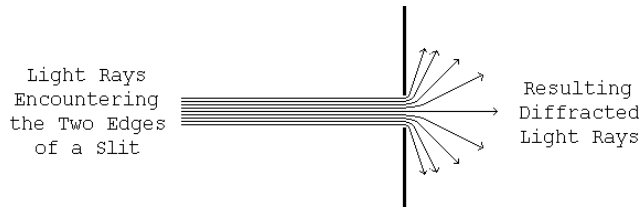

*Figure 4 - Diffraction at a Slit Causing Bending of Light Rays*

In both of these cases, gravitational lensing and slit diffraction, the direction of the wave front is changed because part of the wave front is slowed relative to the rest of it. In the case of gravitational lensing the part of the wave front nearer to the "massive lensing cosmic object" is slowed more. In the case of diffraction at a slit the part of the wave front nearer to the solid, opaque material in which the slit is cut is slowed more.

But, neither of the cases, gravitational lensing and slit diffraction, involves the wave front passing from traveling through one substance to another as in the first illustration, above. The wave front in the gravitational lensing case is traveling only through cosmic space. The wave front in the slit diffraction case is traveling only through air. There is no substance change to produce the slowing. What is it that slows part of their wave fronts thus producing the deflection ?

In the case of gravitational lensing the answer is that the effect is caused by gravitation. There is no other physical effect available. But how does gravitation produce slowing of part of the incoming wave front so as to deflect it ? Gravitation, at least as it is generally known and experienced, causes acceleration, not slowing.

<u>*Electro-Magnetic Field (Light)* and *Gravitational Field (Gravity)*</u>

*1 - Light*

Given two particles [e.g. electrons or protons] that have electric charges, the particles being separated and with the usual electric [Coulomb] force between them, if one of the charged particles is moved the change can produce no effect on the other charge until a time equal to the distance between them divided by the speed of light, $c$, has elapsed.

For that time delay to happen there must be <u>something flowing</u> from one charge to the other at speed $c$ [a fundamental constant of the universe] and each charge must be the source of such a flow.

That electric effect is radially outward from each charge, therefore every charge must be propagating such a flow radially outward in all directions from itself, which flow must be the "electric field".

When such a charge moves with varying velocity it propagates a pattern called electromagnetic field outward into space. Light is that pattern, that field traveling in space. Since light's source is a charged particle that, whether the particle is moving or not, is continuously emitting its radially outward flow that carries the affect of its charge, then light's electromagnetic field is a pattern of variations in that flow due to the charge's varying velocity.

*2 - Gravity*

Given two masses, i.e. particles that have mass [e.g. electrons or protons], being separated and with the usual gravitational force [attraction] between them, if one of the masses is moved the change can produce no effect on the other mass until a time equal to the distance between them divided by the speed of light, $c$, has elapsed.

For that time delay to happen there must be <u>something flowing</u> from one mass to the other at speed $c$ and each particle, each mass must be the source of such a flow.

That gravitational effect is radially outward from each mass, therefore every mass must be propagating such a flow radially outward in all directions from itself, which flow must be the "gravitational field".

*3 - <u>That Flow</u>*

We therefore find that the fundamental particles of atoms, of matter, which have both electric charge and gravitational mass, must have <u>something flowing</u> outward continuously from them and:

- Either the particles have two simultaneous, separate outward flows, one for the effects of electric charge and another for gravitation, or
- They have one common universal outward flow that acts to produce all of the effects: electric and electromagnetic field [light] and gravitational field [gravity].

There is clearly no contest between the alternatives. It would be absurd for there to be two separate, but simultaneous, independent outward flows, for the two different purposes. And, the single outward *Flow* from particles, carrying both the electric and electromagnetic field and the gravitational field, means that gravitational field can have an affect on light, on electro-magnetic field because they both are the same medium – the universal outward *Flow*.

The "gravitational lensing" presented earlier above is experimentally observed gravitational field affecting light.

*GRAVITATIONAL SLOWING / DEFLECTION OF LIGHT*

Because that universal outward *Flow* originates at each particle and flows radially outward in all directions its density or concentration decreases inversely as the square of distance from the source of the *Flow*. At a large distance from the source the wave front of a very small portion of the total spherical outward *Flow* is essentially flat – a "plane *Flow*".

Two such universal *Flow*s directly encountering each other "head on" [flowing exactly toward each other] interfere with each other, that is each slows the *Flow* of the other. The effect is proportional to the density or concentration of each *Flow*.

When two such *Flow*s encounter each other but not directly "head on" then each *Flow* can be analyzed into two components: one directly opposed to the other's *Flow* and one at right angles to that direction.

In "gravitational lensing" gravitational *Flow* produces deflection of the *Flow* that carries light. That deflected *Flow* is the same *Flow* that also simultaneously carries gravitation. Thus the gravitational *Flow* from one mass can also produce deflection of the gravitational *Flow* from another mass.

<u>Therefore, a properly configured material structure can deflect gravitation away from its natural action, reducing the natural gravitation effect on objects that the gravitation would otherwise encounter and attract.</u>

That same effect, on a vastly reduced scale, produces the deflection, the bending of the light direction that is seen in slit diffraction. In the diffraction effect the role of the "massive lensing cosmic object" is performed by the individual atoms making up the opaque thin material in which the slit is cut. That effect shows that the gravitational lensing process, involving immense cosmic masses, can be implemented on Earth on a much smaller scale practical for human use.

## THE ENERGY ASPECT AND THE SOURCE OF THE FLOW

But, changing the "natural gravitation effect" means changing the gravitational potential energy of objects in the changed gravitational field. If the energy is changed where does the difference come from or go to ?

The potential energy for an object of mass, $m$, at a height, $h$, in a gravitational field is truly <u>potential</u>. It is the kinetic energy that the mass <u>would acquire</u> from being accelerated in the gravitational field <u>if it were to fall</u>. The greater the mass, $m$, the greater is the kinetic energy, $\frac{1}{2} \cdot m \cdot v^2$. The greater the distance, $h$, through which the mass would fall the greater the time of the acceleration, the greater the velocity, $v$, achieved, the greater the kinetic energy, $\frac{1}{2} \cdot m \cdot v^2$.

While at rest at height $h$ [as on a shelf] the total mass of the object is the same as its rest mass. The object has no actual "potential energy". It is merely in a situation where it could acquire energy, acquire it by falling in the gravitational field. Falling, the mass of the object increases as its velocity increases, reflecting its gradually acquired kinetic energy.

Since, <u>until it falls, the object does not have the energy that it will acquire when it falls</u> in the gravitational field <u>the energy that it acquires must come from the gravitational field</u>.

The energy of gravitational field is in its *Flow* radially outward from all gravitational masses. The *Flow* is a flow of the potential for energy, realized at any encounter with another gravitational mass

- That *Flow* creates potential energy, <u>creates the situation where kinetic energy could be acquired</u>, at any gravitational mass that it encounters.

- It does so continuously, replenished and replenishing by the on going continuous outward *Flow*.

- It does so continuously, regardless of the number or amount of masses encountered and regardless of their distance from the source of the *Flow*.

- At each encountered mass the amount of the *Flow* varies with the magnitude of its source mass and varies inversely as the square of the distance from it.

But, for there to be a continuous *Flow* outward from each mass particle, each must be a supply, a reservoir, of that medium which is flowing. The original supply of the *Flow* medium, of gravitational potential energy, came into existence at the "Big Bang" the beginning of the universe.

If that immense reservoir of energy could be tapped by tapping some of its appearance in its outward *Flow*, which is the gravitational field, it could be a vast supply of energy cheaply, cleanly, and permanently without [for practical human / Earth purposes] being used up.

Since the original "Big Bang" the outward *Flow* has been very gradually depleting the original supply. That process, an original quantity gradually depleted by flow away of some of the original quantity is an exponential decay process and the rate of the decay is governed by its time constant. In the case of the decay of the universal *Flow*, appearing among other places in the outward *Flow* from every gravitating mass, the time constant is about $\tau = 3.57532 \cdot 10^{17}$ sec ( $\approx 11.3373 \cdot 10^9$ years).

## TAPPING THE ENERGY OF THE GRAVITATIONAL FIELD

The general vertically upward outward *Flow* of gravitational energy can be tapped by deflecting part of a local region's gravitational *Flow* away from its normal vertical direction. Figure 5 below [the slit diffraction figure from earlier above but now rotated 90º] illustrates such deflection using a single slit.

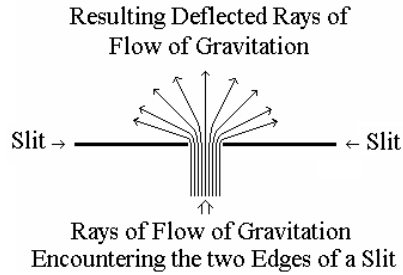

*Figure 5 - Slit Diffraction, the Basic Element of a Gravitation Deflector*

Multiple such slits parallel to each other would spread the deflection left and right in the figure. Additional multiple such slits at right angles to the first ones would spread the deflection over a significant area.

<u>GRAVITATION DEFLECTOR DESIGN</u>

The edges of the slit in the above Figure 5 are actually rows of atoms. A cubic crystal, such as of Silicon, consists of such rows of atoms, multiple rows and rows at right angles, all equally spaced – a naturally occurring configuration of the set of slits required for deflection of gravitation.

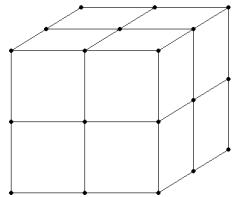

*Figure 6 - A Small Piece of a Cubic Crystal*

The *Flow* from each of the cubic crystal's atoms is radially outward. Therefore its concentration falls off as the square of distance from the atom. The amount of slowing of an incoming gravitational *Flow,* and therefore the amount of its resulting deflection, depends on the relative concentrations of the atoms' *Flow* and the overall gravitational *Flow*.

In the case of diffraction of the *Flow* of light at a slit the concentration of the *Flow* from the atoms of the slit material is comparable to the concentration in the horizontal *Flow* of the light, because it originates from a local source, not from the Earth's immense gravitation.

But for the *Flow* from the atoms of the slit to deflect the much more concentrated vertically upward *Flow* of Earth's gravitation the *Flow* from the atoms of the slit must also be much more concentrated. The only way to achieve that more concentrated *Flow* is a configuration in which the *Flow* of Earth's gravitation is forced to pass much closer to the atoms of the slit so that, per the inverse square variation in the atoms' *Flow*, it will pass through a concentration of the slit atom's *Flow* comparable to the concentration in the Earth's gravitational *Flow*.

The spacing between the edges of the diffracting slit is about $5 \cdot 10^{-6}$ $meters$. The spacing of the atoms at the corners of the "cubes" in a Silicon cubic crystal is $5.4 \cdot 10^{-10}$ $meters$. An inter-atomic spacing of less than $3 \cdot 10^{-19}$ $meters$, much closer than the natural spacing in the Silicon cubic crystal, is required to obtain deflection of a major portion of the incoming Earth's gravitational *Flow*. [2]

Such a close atomic spacing cannot be obtained by directly arranging for, or finding a material that has, such a close atomic spacing. However, that close an atomic spacing can be effectively produced relative to just the vertical *Flow* of gravitation by slightly tilting the Silicon cubic crystal's cubic structure relative to the vertical.

The following Figure 7 illustrates the tilting, schematically and not to scale, and shows how it increases the number of crystal atoms closely encountered by the upward gravitational *Flow*.

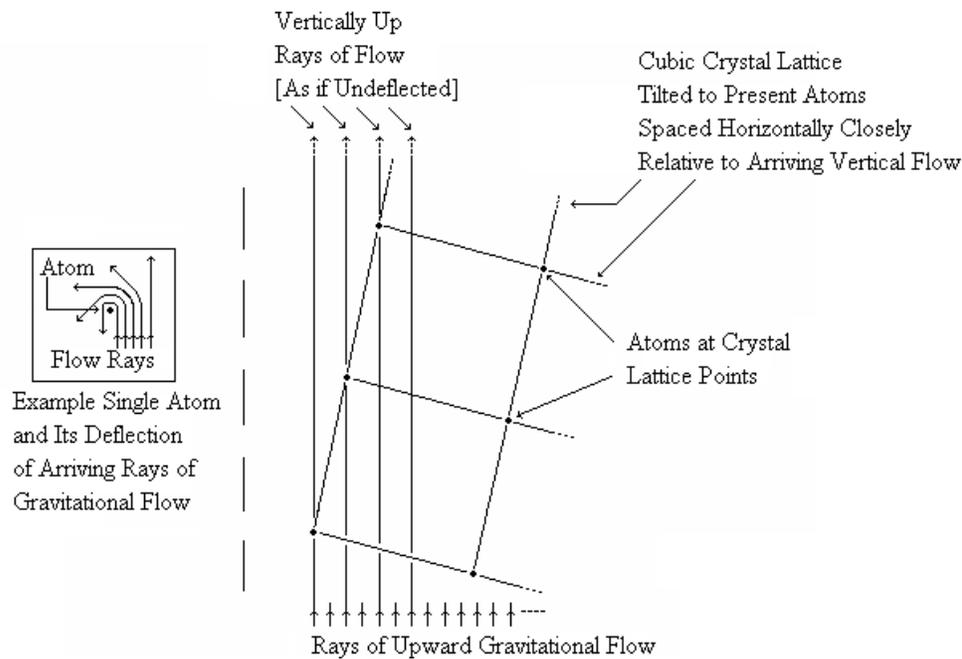

*Figure 7 - Cubic Crystal Lattice Tilted for Effective Gravitational Flow Deflection*

By appropriate tilting of the cubic structure each of its $5.4 \cdot 10^{-10}$ `meters` inter-atomic spaces is effectively sub-divided into $10^{10}$ "sub-spaces" each of them $5.4 \cdot 10^{-20}$ `meters` long and with an atom in each. A `4.5 mm` shim on a `30 cm` diameter Silicon cubic crystal ingot produces such an effect, producing a tilt `tangent = 0.015` for a `tilt angle = 0.86°` that produces the objective effective sub-division of the crystals' natural inter-atomic spacing, a sub-division that acts only on vertical *Flow*, as of gravitation.

Pure, monolithic, Silicon cubic crystals up to `30 cm` in diameter are grown for making the "chips" used in many electronic devices. The gravitation deflector requires a large, thick piece of Silicon cubic crystal rather than the thin wafers sawed from the "mother" crystal for "chip" making.

Per the detailed analysis in the references, The Silicon cubic crystal ingot for the deflector is to be:

- `30 cm` in diameter,
- `50 cm` or more thick,
- with the orientation of the cubic structure marked for proper placement of tilt-generating shims, and
- with the bottom face of the cylinder sawed and polished flat at a single cubic structure plane of atoms.

Mean free path [`MFP`] is the average straight line distance a moving particle travels between encounters with another particle. For atoms in solid matter the mean free path is

$$\text{MFP} = \frac{1}{[\text{Atoms Per Unit Volume}] \cdot [\text{Atom Cross Section Area}]}$$

For the Earth the atoms per unit volume is on the order of

```
Atoms per Unit Volume = 5·10^28 per cubic meter.
```

In the cubic crystal deflector the atomic spacing produced by the tilt is about $10^{-20}$ meters. Each therefore has cross sectional space available to it of that of a circle of that diameter so that for this purpose the atom's cross section area is

$$\text{Atom Cross Section Area} = \pi/4 \cdot [10^{-20}]^2$$
$$= 8 \cdot 10^{-39} \text{ meter}^2$$

For targets as fine as those in the cubic crystal deflector, the mean free path in the Earth's outer layers is, therefore

```
MFP = 2.5·10^9 meters
```

The mean free path in the *50 cm* thick minutely tilted Silicon cubic crystal ingot for intercepting Earth's natural <u>vertically</u> outward gravitation is ½ the *50 cm* thickness of the ingot. The gravitation deflector is about $10^{10}$ times more effective than the natural Earth at intercepting Earth's natural gravitation. However, that effectiveness is only for vertical rays of *Flow*. The Silicon crystal's mean free path for non-vertical *Flow – Flow* already once deflected within the crystal – is that of Earth, $2.5 \cdot 10^9$ *meters*, which takes the once-deflected *Flow* out of the crystal.

The overall deflector consists of:

- A support having a verified perfectly horizontal upper surface for the cubic crystal deflector bottom face to rest upon;

- The Silicon cubic crystal ingot specified above; and

- Precision shims *4.5 mm* thick for producing the tilt of the cubic crystal ingot, the shims located at the mid-point of two adjacent sides of the horizontal plane of the cubic structure as in Figure 8 below.

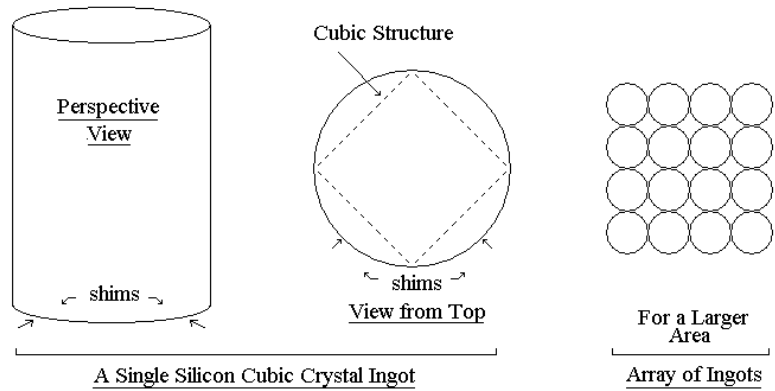

*Figure 8 – The Silicon Cubic Crystals Arrangements*

- For an array of ingots for a larger area than a single ingot can provide, the individual ingots can be machined to fit snugly together. That could be done by machining them to a square cross section or, better, to a hexagonal one.

*PRACTICAL ASPECTS AND DESIGN ENGINEERING*

While the net gravitational field is vertically upward, i.e. radially outward from the Earth's surface, local gravitation is radially outward from each particle of matter. As in Figure 9 below, a mass above the Earth's surface receives rays of gravitational attraction from all over its surrounding surface and the underlying body of the Earth.

The net effect of all of the rays' horizontal components is their cancellation to zero.

The net effect of all of the rays' vertical components is Earth-radially-outward gravitation.

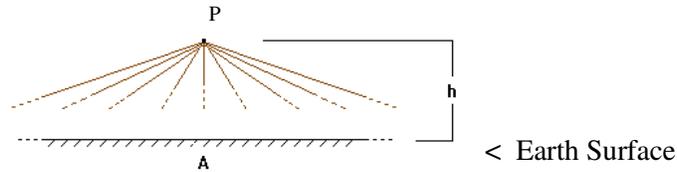 < Earth Surface

*Figure 9 - Rays of Gravitation from the Surroundings*

*1 - Gravitational Ray's Horizontal and Vertical Components.*

One can consider all of the net gravitational effect on objects as being due to the vertical component of all of the myriad rays of gravitational field *Flow* at a wide variety of angles to the horizontal. This "components aspect" is valid because of the "components aspect" appearance in the "Gravitational Lensing" effect on cosmic light.

The various rays of the *Flow* propagation from the individual particles of the gravitating body [for example the Earth] are from each individual particle of it to the selected point [above the gravitating body] on which their action is being evaluated. That is the point `P` in the above Figure 9 directly above the `"A"` at height `h` in the figure.

The Earth's gravitational action along a ray of *Flow* takes place from the Earth's surface to deep within the Earth. The inverse square effect, that the strength of a *Flow* source is reduced as the square of the increase in the radial distance of it from the object acted upon, is exactly offset by that the number of such sources acting [per "ray" so to speak] increases as the square [non-inverse] of that same radial distance. That is, the volume, hence the number, of *Flow* sources for a ray of propagation at the object is contained in a conical volume, symmetrically around the ray with its apex at the object acted upon.

However, because the net gravitational effect is produced only by the vertical component of each ray of *Flow* propagation, the effectiveness of each ray is proportional to the Cosine of the angle between that ray and the perfectly vertical as the angle `θ` in Figure 10 below.

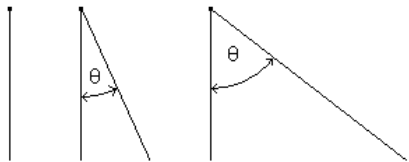

*Figure 10 – The Gravitational Field Ray Angle to the Vertical*

The actual total gravitational action includes all rays from `θ = 0` through to `θ = 90°`. That range would require an infinitely large deflector to act on all such rays, that is the deflector would have to be a disk of infinite radius. For lesser values of the maximum `θ` addressed, the portion of the total gravitation sources included is the integral of `Cos θ·dθ` from `θ = 0 to θ = Chosen Lesser Value`. The integral of the *cosine* is the *sine*. Example lesser portions of the total gravitational action addressed as `θ` varies are presented in the table below.

| θ | Sin θ = Fraction of Total Maximum Gravitational Action |
|---|---|
| 0° | 0.000 |
| 30° | 0.500 |
| 45° | 0.707 |
| 60° | 0.866 |

The gravitational deflector as a disk beneath the *Object* to be levitated must extend horizontally far enough to intercept and deflect the `Chosen Lesser Value` of angle `θ` rays of gravitational wave *Flow* that are able to act on the *Object* of the deflection as depicted in Figure 11 below.

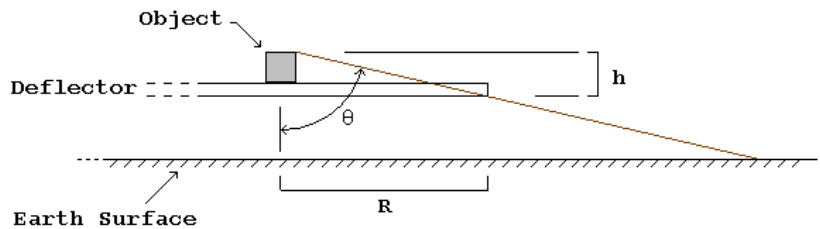

*Figure 11 – Size Requirements for a Disk Shaped Deflector*

For the perfectly vertically traveling rays of gravitation waves the required vertical distance that must be traveled within the cubic crystal is the previously presented `50 cm` and `0` horizontal distance is traversed in so doing. But a ray at angle `θ`, in order to traverse the required 50 cm vertically, must traverse horizontally `50·Tan[θ] cm`, at the same time. For `θ` more than `45°` that can become quite large and the deflector likewise.

Because the deflector disk must extend over a large area to deflect most of the gravitation, an alternative, and better, solution to the problem of rays of gravitation arriving over the range from `θ = 0 to θ = 90°` is to wrap the deflector up the sides of the *Object* to be levitated as shown below.

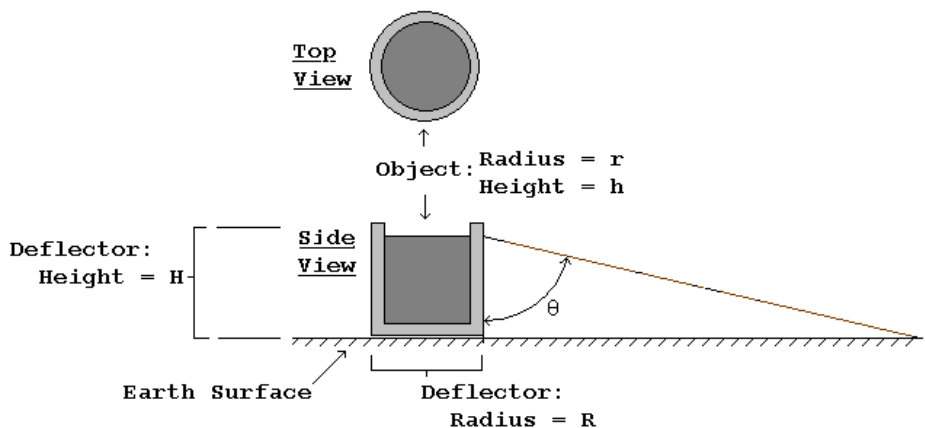

*Figure 12 – A Cup Shaped Gravitation Deflector*

In this configuration the deflector takes up little more space than the *Object* levitated. However, the non-perfectly vertical traveling rays must still travel within the cubic crystal the horizontal distance `50·Tan[θ] cm`. That requires that the horizontal thickness of the vertical sides of the cup-shaped deflector must be of that $50 \cdot Tan[\theta]$ cm thickness.

Because the value of $Sin\ \theta$ and, therefore, the fraction of the total gravitational action, increases relatively little above $\theta = 60°$ whereas the value of $Tan[\theta]$ increases quite rapidly, from $1.7\ to\ \infty$ above $\theta = 60°$ that $\theta = 60°$ is the appropriate value to which to design. The thickness of the "walls" of the "cup" would then be $50 \cdot Tan[60°] = 85$ cm. The deflector would be only slightly larger than the *Object* levitated.

*2 - The Array Structure and Size.*

The Deflector consists of an array of Silicon cubic crystals. The crystals forming the disk-shaped "base" of the "cup" need to be about $0.5\ m$ in height to achieve their maximum deflection effectiveness. Those forming the "sides" of the cup can be the same kind of $0.5\ m$ crystals stacked and aligned vertically.

The crystals can effectively be grown in diameters up to about $30$ cm, however those cylindrical pieces must then be machined down to of hexagonal cross section so that a number of them can fit together with negligible open space between. The hexagonal cross section area would be about $A = 0.06\ m^2$

For an *Object* to be acted upon by the deflector, the object of height, $h$, and diameter, $d$, meters the deflector would have the following parameters for $\theta = 60°$. [The number of crystals must be the integer next higher than the exact calculated number.]

```
Base Disk: Thickness = 1 Crystal Layer = 0.5 m
           Diameter  = d
           Area      = π·d²/4 = 0.785·d²
           Number of crystals = π·d²/4·A
                              = 13.1·d²
Cup Sides:
    Thickness              = 0.85 m
    Outside diameter [OD]  = d + 2·thickness
                           = d + 1.7
    Inside diameter [ID]   = d
    Height                 = h + 2·0.5
                           = h + 1.0
    Height number of Layers = Height/0.5
    Area of Layer = π·[OD²-ID²]/4
    Layer Number of crystals = π·[OD²-ID²]/4·A

Total Number of Crystals:
    Number of Crystals =
      = Base Disk + [Layer Number × Number of Layers]
```

Some examples of these data are presented in the table below.

| d | h | Cup Disk Base | | Cup Sides | | | Total Crystals |
|---|---|---|---|---|---|---|---|
| | | Area | Crystals | Nr of Layers | Area | Crystals | |
| 1 | 1 | 0.785 | 14 | 2 | 4.94 | 99 | 212 |
| 10 | 10 | 78.5 | 1,310 | 20 | 28.97 | 580 | 12910 |

*3 - <u>Calibrating the Individual Silicon Crystals</u>*

The individual crystals making up the deflector cannot be grown exactly identical to each other. In each the orientation of the long axis of the cubic crystal structure may vary minutely from each of the others. That is, it is not certain that each crystal's base is purely a single plane of atoms of the cubic structure and thus is exactly perpendicular to the long axis of the crystal.

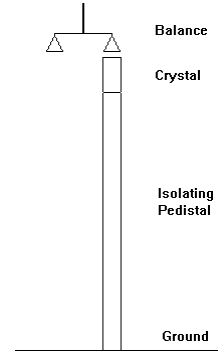

To find the optimum tilt and orientation for a single crystal the tilt must be varied over the range of possibilities while the effect of gravitation from exactly below it is observed on a balance scale. But most of the effect of gravitation on a single crystal is not from exactly below it.

The solution to that problem is to conduct the optimization atop a structure, that relying on the inverse square effect, effectively isolates the crystal from most of the gravitation from surrounding sources except that exactly below it – a high pedestal having a cross section comparable to that of the crystal, as in Figure 13.

*Figure 13*

To conduct that calibration on thousands of crystals should not be necessary if a method can be developed to exactly measure the long axis orientation in any given crystal. The process can then determine the optimum orientation of the crystal tilt relative to the actual long axis of a few cubic crystals being calibrated. That same crystal tilt relative to the actual long axis can then be applied to each of the other crystals.

The long axis orientation problem could also be solved by insuring that the base of each crystal is a single plane of atoms of the cubic structure.

<small>THE AMOUNT OF DEFLECTION</small>

The manner of the deflection is curving of the path of rays of gravitational *Flow* as they pass close to atoms of the deflector with the direction to which curved depending on the relative positions of the ray and an atom and the amount of the curving depending on how close the ray passes to the atom. Because of the range of those variables and their various combinations the "deflection" is essentially a "scattering" in various amounts in various directions, all scattering being away from the perfectly vertical upward which the deflector is designed to solely deflect.

The "scattering" is illustrated two-dimensionally in Figure 14 below. Three dimensionally it can be visualized as that figure viewed from the top while rotated through a full circle.

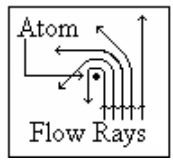

*Figure 14 - Single Atom Deflection of Rays of Gravitational Flow*

The physical example of the "scattering" is the diffraction pattern of light diffracted by a slit. Figure 15, below, presents the diffraction pattern produced by a slit that is `5.4·10⁻⁶ meter` wide with incoming light of wavelength `4.13·10⁻⁷ meter`. The peaks and valleys of the pattern, the interference pattern, are a phenomenon of the light imprint on the *Flow* that carries it. <u>The envelope of the pattern is the relative amounts of the underlying *Flow* carrying the light</u>.

For that reason, while the interference pattern varies according to the wavelength of the light involved, <u>the form of the envelope of that pattern is always the same</u>.

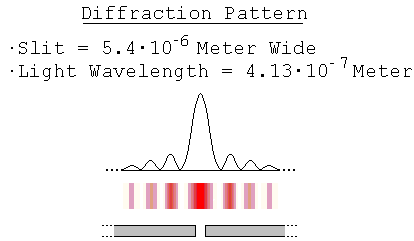

*Figure 15 - A Slit Light Diffraction Pattern*

The *Flow* concentration produced by the two slit edges falls off with distance from the edge inversely as the square of distance from its atoms. The Cauchy-Lorentz Distribution is an inverse square function of its variable. Its Density Function can represent the relative *Flow* intensity pattern produced by the diffraction process by representing the envelope of the diffraction pattern. In Figure 16, the Cauchy-Lorentz distribution is fitted to the diffraction pattern by the appropriate choice of value of its distribution parameter $\gamma$ [Greek *gamma*].

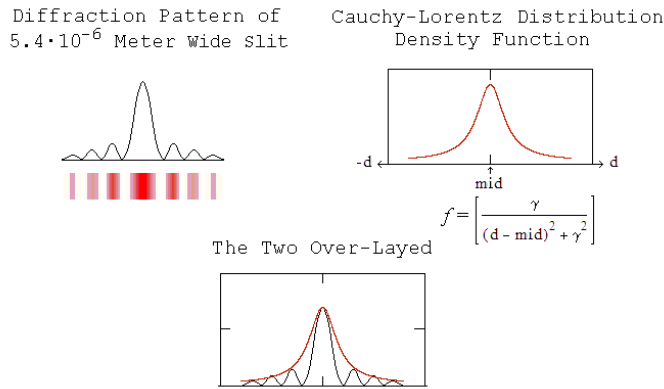

*Figure 16 - The Cauchy-Lorentz Distribution Diffraction Pattern Envelope*

The deflection angle, $\Phi$, is the angle of deflection of the rays to any particular point on the diffraction pattern. That is $\Phi$ is the angle of deflection of the rays directed to that particular point and of intensity per the Cauchy-Lorentz Distribution at that point.

The interest here is not in the location of the light interference maxima and minima, but rather in the deflection angles the diffraction imposes on the *Flow*. However, calculation of the deflection angles to the minima provides a good indication of the amount of *Flow* deflection over the overall diffraction pattern. The table below presents that data for the $5.4 \cdot 10^{-6}$ *meter* wide slit with incoming light of wavelength $4.13 \cdot 10^{-7}$ *meter*. [The minimums are counted outward from the center peak of the diffraction interference pattern].

| Minimum # | $\Phi°$ | Minimum # | $\Phi°$ |
|---|---|---|---|
| 1 | 4.39 | 8 | 37.72 |
| 2 | 8.80 | 9 | 43.50 |
| 3 | 13.26 | 10 | 49.89 |

| 4 | 17.81 | 11 | 57.28 |
|---|---|---|---|
| 5 | 22.48 | 12 | 66.60 |
| 6 | 27.36 | 13 | 83.86 |
| 7 | 32.37 | 14 | $Sin(\Phi) > 1.0$ |

$Sin(\Phi) = n \cdot [\text{ light wavelength } / \text{ slit width }]$, n = 1, 2, ...

*Figure 17 – Table of Diffraction Minimums Deflection Angles*

    The above table demonstrates that the deflection of the *Flow* is at least in amounts up to `90°`.  That deflection may well extend to angles beyond `90°`, but there is no way of determining that from the diffraction pattern.  However, while the light of the diffraction pattern cannot be deflected beyond `90°` in any case because the light cannot penetrate the material containing the slit, the *Flow* readily penetrates and permeates all of material reality.

    The tilt of the cubic crystal structure divides the slit into `10`$^{10}$ sub regions the first and last of which are at the slit's edge and produce the maximum deflection.  The tilt so arranges that ultimately all of the vertical components of the incoming vertical Flow must pass through one of those "at the edge of the slit" regions and must experience maximum deflection.

    The overall average effect is equivalent to every ray's vertical component curving at least *90º* because the crystal tilt causes every ray to pass extremely close to an atom at some point in the crystal, as the extreme rays in the figure below.

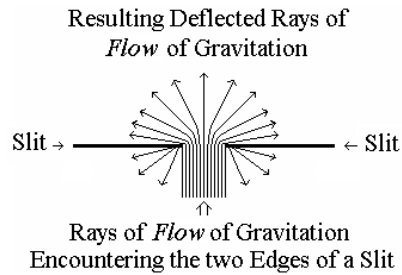

*Figure 18 – Single Slit Gravitation Deflection*

    There does not appear to be any way to analyze, calculate, or evaluate in advance the overall deflection that is achieved other than by actual experiment.  With the overall average effect equivalent to every ray's vertical component curving `90°`, i.e. to the horizontal, the overall total net effect of the vertical components after deflection is zero.  Then the overall amount of deflection is `100%` of the natural un-deflected gravitation reducing the gravitation to essentially zero.

### *GRAVITO – ELECTRIC POWER GENERATION*

    Gravito-electric power generation is similar to hydro-electric power generation in which the energy of water falling in Earth's gravitational field powers water-turbines that drive electric generators.

    In gravito-electric power, depicted schematically in the figure below, a gravitation deflector makes the water in the central region of the mechanism lighter than that in the outer region, which is acted on by natural gravitation.  The lighter reduced gravitation water floats up on the in-flow under it of the heavier natural gravitation water.  The result is continuous circulation of the water, like a continuous waterfall.

    Water turbines like those used in hydro-electric plants can be placed in the gravito-electric continuous water flow to drive electric generators as in hydro-electric plants.

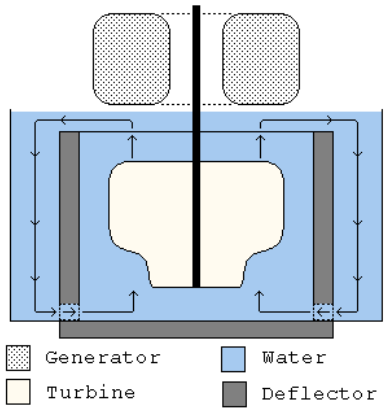

*Figure 19 – Gravito-Electric Power Generation*

## THE DEFLECTION CAUSES A REACTION BACK ON THE DEFLECTOR

Everything in nature is balanced. Nature exhibits a general law of conservation that goes far beyond conservation of energy. For example:

- All positive charge is ultimately, somewhere, balanced by an equal amount of negative charge;

- Gravitational attraction takes place by a mass acting on another mass. The attractive force acting on each is the same in magnitude and opposite in direction; the forces balance;

- The "Big Bang" produced equal amounts of matter and anti-matter;

- For every force there is an equal-but-opposite reaction force;

- Every North magnetic pole is matched by an equal strength South magnetic pole.

As that balance, there is a reaction on the deflection-causing gravitation deflector, a reaction to its deflecting action, a balancing reaction.

The gravitational field *Flow* is an unlimited* capacity to produce acceleration. That is what the outward propagating gravitational field *Flow* does: it accelerates any and every encountered particle of mass no matter how many and no matter where located. But, the amount of gravitational acceleration does not depend on the mass that is accelerated; rather, it is in an amount dependent only on the mass, $M$, of the gravitational *Flow* source and the distance, $d$, from that source to the accelerated mass, which two parameters determine the gravitational field strength at the accelerated mass.

```
Gravitational Acceleration = G·M/d2
```

That *Flow* is what the gravitational deflector deflects.

The associated "force" is that acceleration multiplied by the mass that is accelerated, which can be whatever mass it happens to be. Thus for gravitation the "force" is inconsequential. No "force" is actually there except in our mental concept of the action. It is the acceleration that is the action.

The reaction on the deflector is an "equal but opposite" <u>acceleration of the deflector mechanism away from the source</u> of the before deflection gravitational field *Flow*; that is, it acts in the opposite direction from the direction, toward the source, of the acceleration that undeflected gravitation produces. The deflector experiences that reaction acceleration regardless of the mass of the deflector and no matter what additional mass may be attached to it, which attached mass is accelerated with the deflector.

---

* The gravitational *Flow* is ultimately limited by the general universal decay of all the sources of that *Flow*.

That is because, again, gravitational field *Flow* accelerates any and every encountered particle of mass no matter how many and no matter where located, in amount independent of the mass accelerated, the amount dependent only on the gravitational field strength at the encountered mass.

$$\text{Gravitational Acceleration} = G \cdot M / d^2$$

The direction of the reaction-produced acceleration [repulsion] is the opposite of the direction [attraction] of the before deflection *Flow*-produced acceleration. The magnitude of the reaction acceleration is the same as the magnitude of the deflection, for which see below.

The ultimate result of the deflection action is the combination of reducing the gravitational attractive acceleration of the deflector [and whatever is attached to it] toward the gravitation source plus the introducing of a reactive repulsive acceleration of the deflector [and whatever is attached to it] in the direction away from the gravitation source.

### *THE MECHANISM OF THE ANTI – GRAVITIC ACCELERATION*

One cannot simply rely on that everything in nature is balanced to account for so dramatic an effect as the repulsive acceleration reaction to the deflection of gravitation – an actual anti-gravity. However, the mechanism producing the effect is simple and natural.

First, natural gravitational acceleration is caused by the *Flow* that encounters a particle flow source producing an increase in the ambient *Flow* concentration on the encountered side of the encountered particle. That has the effect of reducing the encountered particle's *Flow* propagation in the direction of that increased *Flow* concentration.

As presented in *The Origin and Its Meaning*[1] that effect creates an imbalance in the propagation, an imbalance that cannot exist. As a result, to correct (or, rather, to prevent) the imbalance, the encountered particle must and does take on increments of velocity in the direction of the increased *Flow* concentration, a resulting acceleration, a gravitational acceleration toward the direction of the increased Flow concentration, which is toward the gravitating source of the *Flow*.

Second, in the case of the deflector, the components of the incoming vertical gravitational field *Flow* that are curved away from the vertical by the deflector's atom's own *Flow* are by virtue of that deflection directed over the side of the atom opposite that facing the source of the gravitation as depicted schematically in Figure 15, above

That increases the *Flow* concentration on that side of the atom. Just as with natural gravitation, that has the effect of reducing the encountered particle's propagation in the vertically upward direction that of the increased *Flow* concentration. That is, the presence on one side of a *Flow* source particle of ambient medium *Flow*, the *Flow* not departing, that is not directed outward away from the particle as is the case for particles' natural propagation, produces the same effect as does natural gravitation, the effect being the same whether the non-departing *Flow* is incoming natural gravitation or deflected gravitational *Flow* passing over.

"Vertical" refers to the direction directly away from the gravitating source. The various individual scattered ray deflections all are a combination of a horizontal component and a vertical component, each in various amounts for various rays. The horizontal components cancel out to null. The vertical components total effect differs from the incoming pre-deflection rays total effect and that difference is the overall amount, or magnitude, of effective deflection.

If every ray's vertical component were curved exactly 90º, i.e. to the horizontal, the total effect of the vertical components after deflection would be zero. Then the overall amount of deflection would be 100% of the natural un-deflected gravitation and the reaction would be acceleration equal in magnitude to the natural un-deflected gravitation but directed away from the source.

The deflection process occurs throughout the length of each deflector crystal. Some rays of gravitational *Flow* are deflected by the first row of atoms of the deflector. Others are deflected by the second row, others the third, and so on. The total deflection is essentially spread linearly uniformly over all of the length of the deflecting crystal.

For the example of every ray's vertical component curved exactly `90°`, i.e. to the horizontal, that would happen linearly uniformly along the crystal length. The result would be that the natural gravitational attraction on the deflector itself would be reduced to `50%` of normal. At the same time the reaction repulsion magnitude would be `100%` of the natural gravitational attraction because of the overall `100%` deflection. The combined effect would be a net repulsive acceleration of magnitude `50%` of the natural pre-deflection attraction.

## *A GRAVITATION DEFLECTOR SPACECRAFT DEEP SPACE DRIVE*

A spacecraft gravitation deflector drive would be a deflector in cup form, mounted on the rear of the spacecraft and extending the spacecraft's full length to the nose, as in Figure 20 below, with engineered arrangements for varying the amount of deflection.

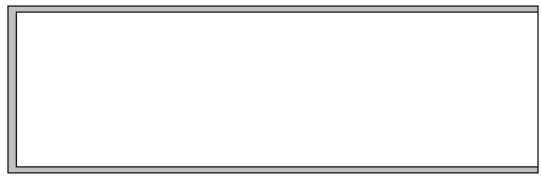

*Figure 20 – A Gravitation Deflector Driven Spacecraft*

This configuration would satisfy a number of functions. The deflector would provide [all without use of fuel]:

- Launching of the spacecraft vertically upward at an upward acceleration of approximately one-half of local natural gravitation, for Earth an acceleration of about `16.1 ft/`$_s$`2`;

- Landing and re-launching of the spacecraft at any gravitating body such as the Moon or Mars;

- Deep space transit propulsion between gravitating bodies;

- Protection from deep space radiation and cosmic ray particles by virtue of the `½` to almost `1` meter thickness of the Silicon deflector;

- A gravity environment within the spacecraft of zero natural gravitation plus an artificial gravitation due to the acceleration of the ship in whatever amount that it is at any particular time [taking "down" as toward the deflector end of the ship].

The engineered arrangements for varying the amount of deflection so as to vary the acceleration would be means of controlled changing of the orientation of selected portions of the Silicon cubic crystals so that they fail to provide the comprehensive deflection of all incoming vertical rays of *Flow*. The engineered arrangements for varying the direction or orientation of the spacecraft would be a 3-axis system of angular momentum wheels

For a spaceship in free space the gravitational *Flow* environment is different from on Earth. In the case of only one gravitation source near enough to be of any important effect and that sole source at a considerable distance from the spaceship, the gravitational *Flow* from that source to the spaceship is essentially all parallel rays. Departing such a source after launch from it requires simply aiming the stern of the ship toward that source. Controlled landing on it requires simply aiming the stern of the ship toward that source and controlling the acceleration by varying the deflection.

In general, however, in deep inter-planetary space gravitation is present albeit fairly weekly because of inverse square reduction of intensity, and it is present in various amounts with attraction toward various differently located sources. As with the sailing navigation using the wind as in earlier centuries, spaceship travel within the Solar System may require techniques analogous to: sail craft's tacking on various headings, "crabbing" into partial "cross wind" as aircraft do, and in general going "where the winds permit". In the spacecraft case the "winds" are the various direction gravitational *Flows* available from which to generate acceleration and to which the spacecraft is subject to attraction.

Solar System navigation is further complicated by the destination's continuous motion. The navigation must be toward where the destination will be upon spacecraft arrival at it as compared to where the destination currently is.

For inter-stellar navigation there is the possibility of near light speed travel. The deflector could provide continuous, fuel-less acceleration to the spacecraft throughout its trip. The continuous acceleration would accelerate the craft during the first part and, with the craft re-oriented using the 3-axis system of angular momentum wheels, decelerate the craft for approach to the destination.

Because the acceleration is independent of the mass of the spacecraft it could be quite large and able to carry everything needed for an extended trip and for survival at the destination. The relatively narrow form of the spacecraft is chosen in Figure 20 because it provides better shielding against deep space radiation and cosmic rays. A different shape might be chosen for a quite large spacecraft: a single storey flat disk or a multi-storey and wide cylinder.

*A GRAVITATION DEFLECTOR PLANET SURFACE FLYING VEHICLE*

A gravitation deflector flying vehicle would be a deflector in cup form, underneath the payload compartment of the vehicle as in Figure 21 below.

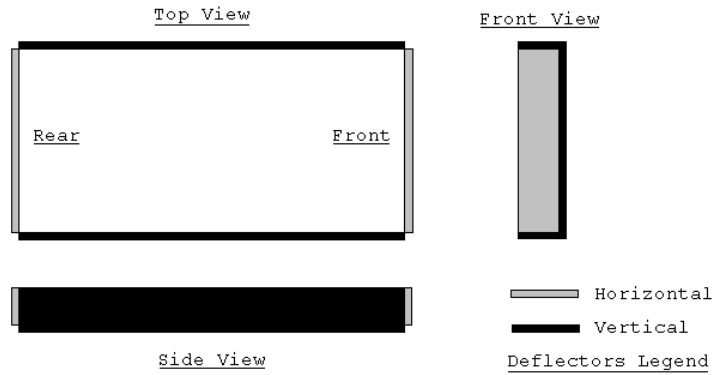

*Figure 21 –A Gravitation Deflector Flying Vehicle*

The flying vehicle differs from the form for a spacecraft in:

- not needing to provide protection from dangerous radiation,

- needing only modest acceleration capability vertically upward beyond sufficient to maintain its constant altitude levitation,

- needing means to generate horizontal acceleration while maintaining vertical levitation.

This deflector configuration [all without use of fuel]:

- Provides controlled vehicle levitation for take-off, landing, and travel,

- Provides controlled horizontal propulsive acceleration and "braking",

- But there is the problem of sufficient gravity for the passengers.

The vertical acting deflectors cannot provide artificial gravity by virtue of vertical acceleration because the vertical acceleration is controlled to only maintain levitation at a given altitude except for take-off and landing. However, maintaining levitation requires significantly less than 100% vertical deflection. If, for example, levitation required only 50% vertical deflection then the gravitation within the vehicle would be the remaining undeflected 50% of natural gravitation.

The present task, then, is research and development to better optimize the designs so that practical implementation can begin.

See the references, below, for the detailed development of mass, field, the *Flow*, gravitation, deflection, the exponential decay and the time constant.

*References*

[1]  R. Ellman, *The Origin and Its Meaning*, The-Origin Foundation, Inc.,
     http://www.The-Origin.org, 1996.

[2]  R. Ellman, *Gravitics – The Physics of the Behavior and Control of
     Gravitation*, The-Origin Foundation, Inc. http://www.The-Origin.org, 2008.